\documentclass[]{spie}  

\usepackage{amsmath,amsfonts,amssymb}
\usepackage{graphicx}
\usepackage[colorlinks=true, allcolors=blue]{hyperref}

\title{Repurposing ROACH-1 boards for prototyping of readout systems for optical-NIR MKIDs}

\author[a]{Ois\'{i}n Creaner}
\author[c,a]{Colm Bracken}
\author[a,b]{Jack Piercy}
\author[a,b]{Gerhard Ulbricht}
\author[a,b]{Eoin Baldwin}
\author[a,b]{Mario De Lucia}
\author[a,b]{Tom Ray}
\affil[a]{Dublin Institute for Advanced Studies, 31 Fitzwilliam Place, D02 XF86, Dublin 2, Ireland}
\affil[b]{Trinity College Dublin, College Green, D02 PN40, Dublin 2, Ireland}
\affil[c]{Maynooth University, Maynooth, Co Kildare, Ireland}

\authorinfo{Further author information: (Send correspondence to O.C.)\\E-mail: creanero@cp.dias.ie, creanero@gmail.com}

\begin{document} 
\maketitle

\begin{abstract}
Microwave Kinetic Inductance Detectors (MKIDs) are cryogenic photon detectors and are attractive because they permit simultaneous time, energy and spatial resolution of faint astronomical sources. We present a cost-effective alternative to dedicated (e.g. analogue) electronics for prototyping readout of single-pixel Optical/NIR MKIDs by repurposing existing and well-known ROACH-1 boards. We also present a pipeline that modernises previously-developed software and data frameworks to allow for extensiblity to new applications and portability to new hardware (e.g. Xilinx ZCU111 or 2x2 RFSoC boards). 
\end{abstract}


\section{INTRODUCTION}
\label{sec:intro}  

Microwave Kinetic Inductance Detectors (MKIDs) are cryogenic photon detectors which consist of superconducting LC resonators and which operate by the kinetic inductance effect \cite{Day2003}.  They are read by observing changes in the frequency of this resonance\cite{Day2003,mkidsapplications}.  This change is proportional to the energy of the incident photon\cite{Day2003}.

They are thus attractive to astronomers \cite{mkidsapplications} because they permit simultaneous temporal resolution and energy resolution.  Time resolution for MKIDs are typically of the order of 1-2\textmu{}s\cite{strader2016digitial,fruitwala2020second}.  Energy resolution of up to $\frac{E}{\Delta{}E}\approx{}55$\cite{deVissersphononiccrystal} has been achieved. Additionally, by creating an array of MKIDs each with slightly different resonant frequencies, multiple MKIDs can be readout on the same feedline through frequency-domain multiplexing\cite{baldwin2020frequency,dibert2022development,baldwin2021}.

\begin{figure} [ht]
   \begin{center}
   \begin{tabular}{c} 
   \includegraphics[height=10cm]{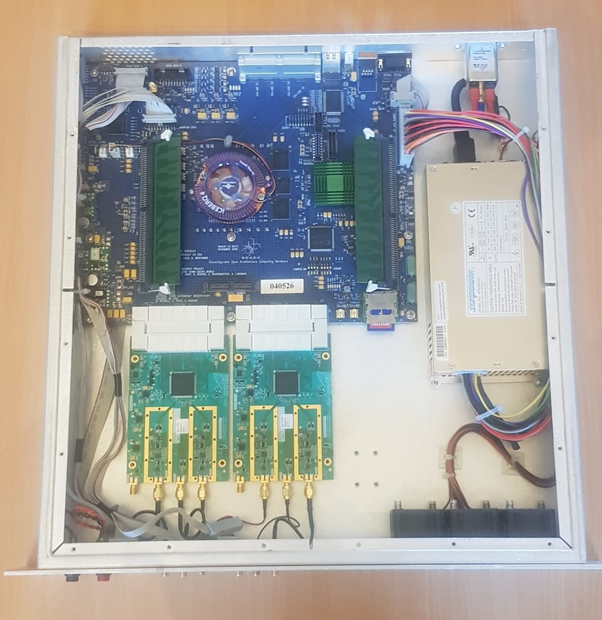}
   \end{tabular}
   \end{center}
   \caption[roach] 
   { \label{fig:roach} 
Image of the ROACH-1 board used in this project.  Visible are the Analogue-to-Digital Converter (ADC) and Digital-to-Analogue Converter (DAC) boards (bottom left of the image) as well as the Field Programmable Gate Array (FPGA) board (top left of image).  Adapted from Ref.~\citenum{eointhesis}}
   \end{figure}

We present a report on an aspect of ongoing project \cite{eointhesis,mariothesis} to develop and test new MKID designs as described elsewhere in these proceedings by Ref.~\citenum{ulbricht2022,delucia2022}.  This paper describes the work undertaken to modernise and streamline the software and firmware system for the Re-configurable Open Access Computer Hardware (ROACH-1)\cite{casperhickish,roachpage} board shown in Figure \ref{fig:roach} used to iteratively test prototype MKIDs.

Our challenge is to identify continued uses that can be made of existing equipment such as the ROACH-1.  This will allow us to address constraints such as budgetary and human limitations, and careful design will allow the lessons learned here to be expanded to future implementations (e.g. in arrays.)  On the other hand that these uses do not negatively impact future expansions by imposing design constraints on systems which will be reused and expanded.

\section{PURPOSE}
\label{purpose}
\begin{figure} [ht]
   \begin{center}
   \begin{tabular}{c} 
   \includegraphics[height=5cm]{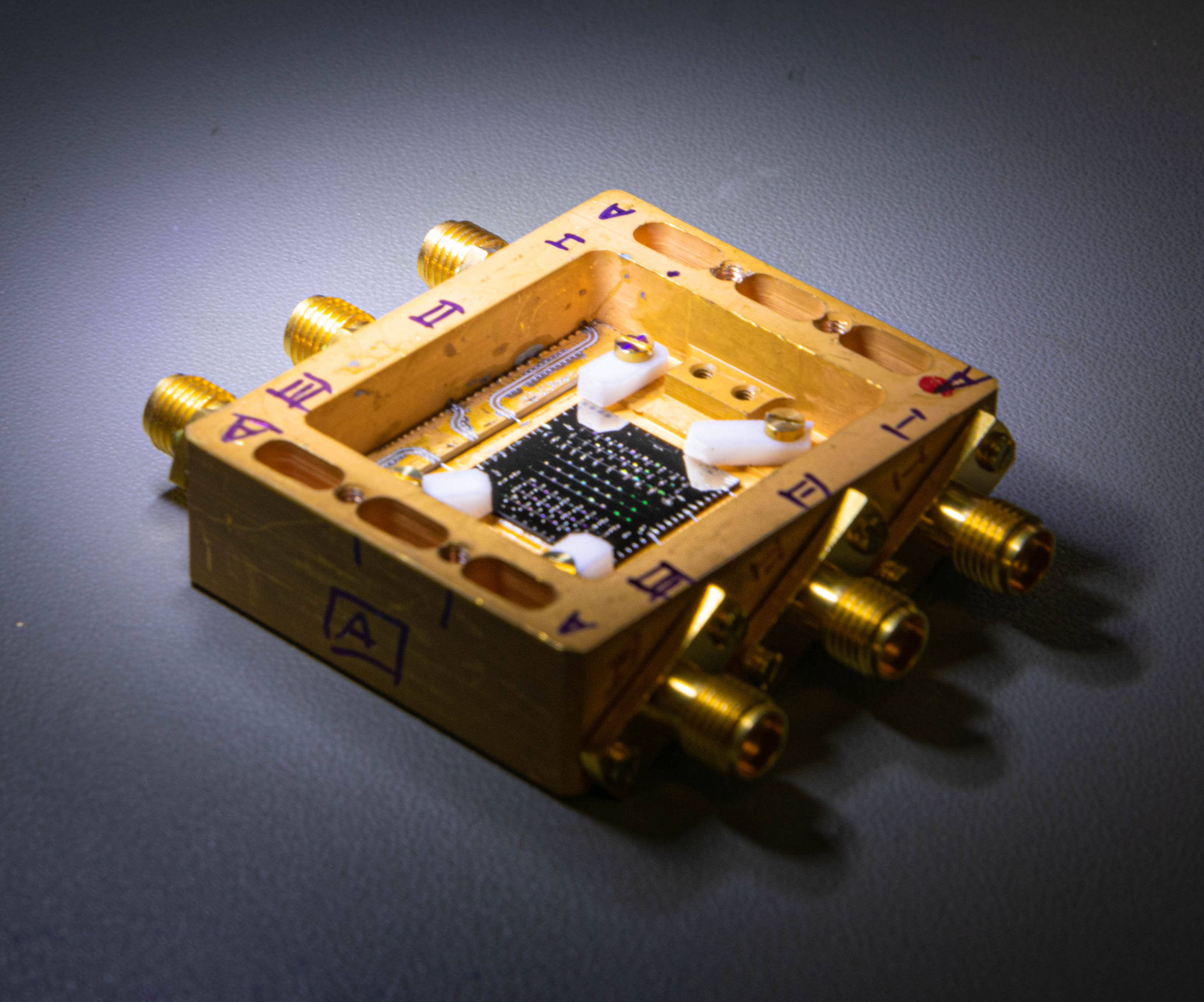}
   \end{tabular}
   \end{center}
   \caption[sample] 
   { \label{fig:sample_box} 
Image of an MKID prototype array produced in the course of this project, held in a sample box to allow for testing. Adapted from Ref.~\citenum{mariothesis}}
   \end{figure}

Our research focuses on the development of single MKID pixels in prototype arrays like the one shown in Figure \ref{fig:sample_box} for prototyping various design parameters and fabrication procedures\cite{ulbricht2022,delucia2022}. For example, by varying the relative thickness of TiN/Ti/TiN multilayers, it is possible to select a critical temperature intermediate between that of Ti and TiN\cite{vissers2013proximity,delucia2022}.  Varying fabrication techniques can also lead to differences in noise characteristics or other important parameters\cite{ulbricht2022}. Therefore it is essential that a system that permits rapid testing be developed to allow these parameters to be varied in an iterative development process.

In this context, the multiplexing capabilities of MKIDs are not necessary to exploit at this stage.  Therefore the difference in readout capacity between the venerable ROACH-1 board, which can be expected to support the readout of $\sim$256 MKIDs\cite{Mazin_2013}, and later models such as the ZCU111 ($\sim$4000) or RFSoC 2x2 ($\sim$2000) \cite{baldwin2021} is irrelevant for our purposes.  

Because the firmware for these boards shares the Collaboration for Astronomy Signal Processing and Electronics Research (CASPER) approach of using Simulink blocks which carry out high-level functions and abstract the low-level functions from the developer\cite{casperpage}, workflows developed for the older equipment should be relatively easy to convert to the newer systems at a later date.

Existing software, developed from an earlier version by Ref.~\citenum{sdrrepo}, sets up the readout of a single MKID pixel measurement by defining resonant frequency, resonator drive power and IQ loop center.  It does this by providing an interface that allows the user to: 
\begin{itemize}
  \item Perform a frequency sweep around an user-defined frequency (i.e. the predicted resonant frequency) to obtain these required parameters
  \item Adjust the ADC input attenuation to make use of the full ADC dynamic range
  \item Monitor a single MKID pixel's 2-dimensional IQ signal with a 1 MHz sampling rate
  \item Calculate a 1-dimensional phase signal from the IQ value and trigger on it
  \item Save a defined number of measurements to disk each time the trigger was reached
\end{itemize}

\subsection{CHALLENGE}
\label{challenge}
Within the overall scope of this project, this paper addresses a specific challenge.  The efficient development of streamlined software to allow the ROACH-1 board to continue to be used for prototype testing in the face of a number of practical constraints.

The readout software for this project was forked from Ref.~\citenum{sdrrepo} at an early stage, and heavily adapted for our purposes.  It consists of a series of shell scripts to configure the hardware environment and python scripts that interface with this environment and allow for the programming of the board.  Developed originally for the ARCONS instrument\cite{Mazin_2013}, it includes features which are irrelevant for our prototyping purpose, such as multi-channel controls. In addition, some of the terms of art associated with ARCONS may not be obvious to a new user.  Finally, because the code was forked and then heavily adapted, it is not possible to benefit from ongoing updates to the original work.  Our recent effort has been to refactor this code to streamline it for our purposes and relieve some of the technical debt that has built up.

Because our work focuses on the prototyping aspect of MKIDs, we have frequent need to carry out post-observation analysis of our data.  For example, when identifying noise characteristics of the data\cite{ulbricht2022}, there is need for post-processing.  An existing code base for this process was available from Ref.~\citenum{ulbricht2015highly}, which was originally developed for X-ray imaging KIDs, but which could be readily adapted to our project.  This software was written in IDL (Interactive Data Language)\cite{IDL_Source}, which is a proprietary language that requires expensive licences to compile. Additionally, the input and output data formats for this software are rigid binary data structures which are designed to be storage-efficient for their original purposes, but limit the scope of their extensiblity and which require bespoke reading and writing code.  For these reasons, it was decided to convert this software to Python \cite{van2007python} and the data formats to JavaScript Object Notation (JSON) format\cite{crockford2006application}.

\subsection{HUMAN FACTOR}
\label{human}
In an academic environment such as is the case with our project, one must account for periodic changes in team composition.  Ideally, this requires a streamlined onboarding process and avoidance of ``Single point of failure'' team members: i.e. individuals who are the only one familiar with certain laboratory procedures.

To mitigate these factors, it is essential that good documentation be developed alongside software systems.  These documents take two main forms: User-level documentation which describes \textit{what} needs to be done to use the system, and Developer-level documentation which describes the software itself, and which is essential to understand \textit{how} the software operates, which is essential for maintenance and future development.

In addition, introduction of new users to the group is simplified by the development of a streamlined front-end which is easily understood and which has design affordances which guide the user towards the intended use.

\section{DESIGN APPROACH}
\label{approach}
In the refactoring and translation of the existing code base, the approach was to focus on five key factors: Usability, Transparency, Portability, Extensiblity and Modularity.

\textbf{Usability} has been emphasised by identifying actions and interactions that the users need, and simplifying how to access them.  Reasonable default values have been set for rarely changed settings, and interdependent variables where the user may wish to vary either term to produce an effect on the other have been presented as linked variables on the front-end.  Procedural steps where the user was previously required to edit the code have been reconfigured to be available as part of the front end and their purposes clearly marked.

\textbf{Transparency} has been improved by ensuring that the units for all variables are clearly displayed with standard SI prefixes. Terms-of-art specific to the original use-case of the software have been replaced with terms which can be more readily understood by the current users.  Space-efficient binary data formats which require bespoke read-write operations have been replaced with human-readable data formats which lend themselves well to modern data analytics techniques.

\textbf{Portability} was included in the design where possible by migrating to more modern and open-source languages, versions and platforms.  This approach extended to both software and data aspects of the project.

\textbf{Extensiblity} was maintained by use of the ``for-but-not-with'' approach: for example while the current application does not require multiplexible readout for multiple MKIDs, we have retained the capacity for multi-channel readout by storing the data as iterable collections such as lists. In the prototyping stage, these collections will typically be of size 1, while in later applications these collections can be made larger without necessitating any code changes.  

\textbf{Modularity} has been worked into the existing codebase where possible to allow updates and maintenance to take place without extensive rewriting.  Additionally, modular data design enabled by the migration to \texttt{.json} format instead of fixed-format binary data allows for new information to be recorded in the same format without interrupting older functionality.

Code changes have taken place in an iterative development process.  A short cycle of design-develop-test allows for improvements to be made without impacting the physical testing processes the software is intended to support.  Where possible, isolated unit tests precede integration testing of new elements of the system.  Once the developer has tested the system, hands-on user testing is carried out to ensure that the needs of the users have been met adequately.  Other factors which have been considered include adding error-checking and graceful failure states to the system to prevent unexpected behaviour or crashes. By following these design approaches, the system is more able to tolerate changes in team composition.  

\section{PROGRESS}
\label{progress}
This paper presents a work-in-progress.  Thus far we have made significant progress on streamlining the front-end user experience of the readout software.  Additional back-end changes have been made to modularise the software system and smooth out the workflow and dataflow.   We have also implemented changes in the data format for output from the readout software for the ROACH board and input to the analysis package. Finally, we have begun a progress to translate the existing post observation analysis software from IDL to Python.   

   \begin{figure} [ht]
   \begin{center}
   \begin{tabular}{c} 
   \includegraphics[height=12cm]{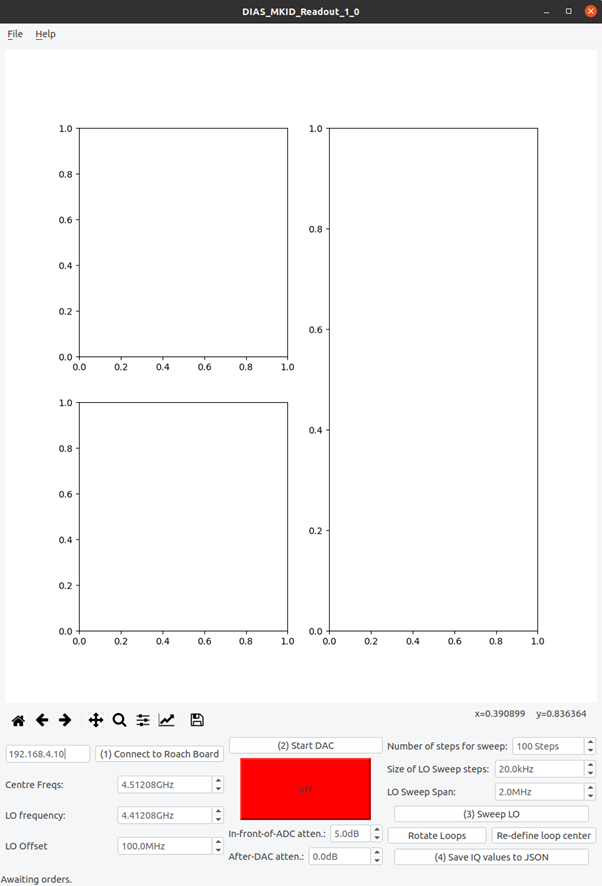}
   \end{tabular}
   \end{center}
   \caption[frontend] 
   { \label{fig:frontend} 
A screenshot of the new front-end for ROACH board readout as adapted for this project. Visible are the added, linked SI SpinBoxes and streamlined UI.  The chart sections are currently blank, indicating that this screenshot was taken before connecting to the board and running an analysis.  Once data is gathered, the axes are adjusted automatically to suit the data and labelled appropriately.}
   \end{figure}

The changes to the front end have emphasised usability and readability, and an illustrative example is shown in Figure \ref{fig:frontend}.  A default value for the ROACH board IP address used in our project was added, and no longer must be specified at run time, though the user can still specify a different value at runtime.  Because we use a single MKID, and thus a single frequency is needed, the input for frequency has been changed to a specially developed SI SpinBox module. This constrains input to numeric values that can include decimal, floating point or SI prefix notations. This module can be seamlessly replaced with an existing input module for a list of frequencies if array readout is needed. In prototype testing, it is necessary to offset the Local Oscillator (LO) frequency from the resonant frequency by a user-defined amount.  Where previously this was set directly by the user, either the offset value or the LO frequency can be set directly through a set of linked SpinBoxes, with either adjusting the other appropriately.  Similarly, when it is necessary to sweep the LO through a series of frequencies for testing, the user can specify the full span of the sweep, the step-size or the number of steps through a set of linked SpinBoxes.  The labels of all of the numerical variables have been adjusted to include their units with appropriate SI prefixes.  The back-end of this software has been streamlined in a number of ways, including by removing of unnecessary write-to-disc steps, connecting scripts which necessarily depend on one another and correcting calls which write to the board and wait for a response to do so in the correct order.  The software is available from Ref.~\citenum{mkidssdrrepo}.

   \begin{figure} [ht]
   \begin{center}
   \begin{tabular}{c} 
   \includegraphics[height=5cm]{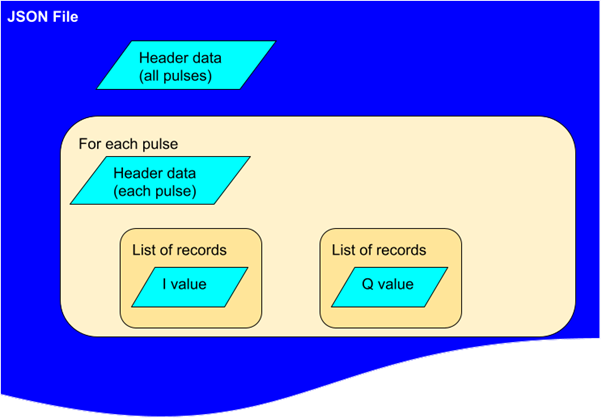}
   \end{tabular}
   \end{center}
   \caption[json] 
   { \label{fig:json} 
Design diagram of the JSON structure used for input/output of data in the new elements of this project}
   \end{figure}
   
We have developed a new data format using the JSON notation \cite{crockford2006application} as illustrated in Figure \ref{fig:json}.  This structure stores information about the experimental setup in a header, and then a series of headers for each result of this observation in a nested structure of keyword:value pairs.  The example shown in Figure \ref{fig:json} shows the structure for storing the results of a series of measurements of pulses in I/Q space which are observed when a photon strikes the MKID under observation.  Each pulse can be examined separately or the data aggregated as needed in the analytic software.  This structure allows for extensibility, as additional information such as frequency inputs during a frequency sweep or timestamp values can be included in this structure.  The analysis system can later open the file and extract the data into a Python dictionary which the user can further analyse.

Translation work has begun to redevelop the analysis software from Ref.~\citenum{ulbricht2015highly} into a suitable Python package.  The challenges in this phase of the project have included 
\begin{itemize}
  \item Replacement of the existing data structures with more flexible Python equivalents
  \item Identifying corresponding Python libraries for the IDL analysis procedures  
  \item Interpreting the relations between data in large functions, especially given IDL's singular namespace
\end{itemize}
These have been addressed by close consultation with the original developers and an iterative development process which allows for testing of the new version in stages.  The code for this software is available from Ref.~\citenum{idltranslate}

\acknowledgments 
 
This project has received funding under Science Foundation Ireland Grant number 15/IA/2880.

This project makes use of Python libraries Corr\cite{corrrepo}, Matplotlib\cite{Hunter:2007}, PyQt4\cite{PyQT4}, Numpy\cite{harris2020array} and Pandas\cite{reback2020pandas,mckinney-proc-scipy-2010}.

Samples shown in Figure \ref{fig:sample_box} and used for testing the presented firmware have been fabricated by the Process \& Product Development team at Tyndall National Institute, Cork, Ireland.

\bibliography{report} 
\bibliographystyle{spiebib} 

\end{document}